\documentclass[final,5p,times,twocolumn]{elsarticle}

\usepackage{graphicx}
\usepackage{fixltx2e}
\usepackage{amssymb}

\journal{Physics Letters B}

\begin{document}

\begin{frontmatter}



\title{Three- and four-nucleon absorption processes observed in the $K^-$-$^4$He reaction at rest}

\author[RIKEN]{T. Suzuki \corref{cor1}\fnref{fn1}}
\ead{tsuzuki@phys.s.u-tokyo.ac.jp}
\fntext[fn1]{Current address : Department of Physics, The University of Tokyo, Tokyo 113-0033, Japan}
\cortext[cor1]{Corresponding author}

\author[SNU]{H. Bhang}

\author[TUS]{J. Chiba}

\author[SNU]{S. Choi}

\author[TITEC]{Y. Fukuda}

\author[TUS]{T. Hanaki}

\author[UTS]{R.S. Hayano}

\author[RIKEN]{M. Iio}

\author[UTS]{T. Ishikawa}

\author[KEK]{S. Ishimoto}  

\author[SMI]{\\ T. Ishiwatari}

\author[RIKEN]{K. Itahashi}

\author[KEK]{M. Iwai}

\author[RIKEN,TITEC]{M. Iwasaki}

\author[TUM,SMI]{P. Kienle}

\author[SNU]{J.H. Kim\fnref{fn2}}
\fntext[fn2]{Current address : Korea Research Institute of Standard and Science, Daejeon 305-600, Republic of Korea}

\author[RIKEN]{Y. Matsuda\fnref{fn3}}
\fntext[fn3]{Current address : Department of Physics, The University of Tokyo, Tokyo 153-8902, Japan}

\author[RIKEN]{H. Ohnishi}

\author[RIKEN]{S. Okada\fnref{fn4}}
\fntext[fn4]{Current address : INFN Laboratori Nazionali di Frascati, I-00044 Frascati, Italy}

\author[RIKEN]{\\ H. Outa}

\author[TITEC]{M. Sato\fnref{fn1}}

\author[KEK]{S. Suzuki}

\author[RIKEN]{D. Tomono}

\author[SMI]{E. Widmann}

\author[UTS,RIKEN]{T. Yamazaki}

\author[SNU]{H. Yim\fnref{fn5}}
\fntext[fn5]{Current address : Korea Institute of Radiological and Medical Sciences, Seoul 139-706, Republic of Korea }

\address[RIKEN]{RIKEN Nishina Center, RIKEN, Saitama 351-0198, Japan}
\address[SNU]{School of Physis, Seoul National University, Seoul 151-742, Republic of Korea}
\address[TUS]{Department of Physis, Tokyo University of Science, Chiba 278-8510, Japan}
\address[TITEC]{Department of Physics, Tokyo Institute of Technology, Tokyo 152-8551, Japan}
\address[UTS]{Department of Physics, The University of Tokyo, Tokyo 113-0033, Japan}
\address[KEK]{High Energy Accelerator Research Organization (KEK), Ibaraki 305-0801, Japan}
\address[SMI]{Stefan Meyer Institut f\"{u}r Subatomare Physik, A-1090 Wien, Austria}
\address[TUM]{Excellence Cluster Universe, Technische Universit\"{a}t M\"{u}nchen, D-85748 Garching, Germany}

\begin{abstract}
Correlations of back-to-back coincident $\Lambda d$ and $\Lambda t$ pairs from the stopped $K^-$ reaction on $^4$He had been investigated, thereby $\Lambda d$ and $\Lambda t$ branches of non-mesonic three- and four-nucleon absorption processes of antikaon at rest were identified as well-separable processes, respectively. The branching ratio of the three-nucleon process, ($^4$He-$K^-$)$_{atomic} \rightarrow \Lambda d ``n"$, is estimated to be $( 0.9 \pm 0.1 (stat) \pm 0.2 (syst)) \times 10^{-3}$ from the normalized $\Lambda d$ spectrum in a $\Lambda d n$ final state, while the fraction of the four-nucleon process, ($^4$He-$K^-$)$_{atomic} \rightarrow \Lambda t$,  is obtained to be $( 3.1 \pm 0.4 (stat) \pm 0.5 (syst)) \times 10^{-4}$ per stopped $K^-$.
\end{abstract}


\end{frontmatter}


\section{Introduction}\label{INTRO}
Until very recently, information on the multinucleonic $K^-$ meson absorption at rest in various nuclei had been mostly confined to measurements of the total multinucleonic capture rate~\cite{Katz, nonmesonic-on-Carbon}. One of only a few exceptions is a full kinematic analysis of the events resulting from $\bar{K}-NN$ interaction performed in deuterium, leading to a determination of the branching ratios of all the two-body final states~\cite{Burnstein}.
A similar analysis in heavier nuclei is difficult due to the large number of underconstrained final states, and only a crude kinematic analysis to the $\Lambda dn$ final state with the $\Sigma^0 dn$ contaminant was performed, which led the first observation of the two-Nucleon Absorption (2NA) process in $^4$He~\cite{Roosen},
 \begin{equation}
 (K^- - {}^4\textrm{He})_{atomic} \rightarrow \Lambda + n + ``d":\textrm{2NA},
 \end{equation}
where $``d"$ is a deueron as the reaction spectator. The observed branching fraction of the reaction, $3.5\pm 0.2 \%$, is significantly smaller than the reported total fraction of 11.7$\pm 2.4\%$ to $\Lambda / \Sigma^0 +pnn/dn/t$ final states~\cite{Katz}, but further items of an account has never been given anywhere until the present. Accordingly, the dynamical properties or even the existence of three- and four-nucleon absorption processes which had been already established to $\pi$ mesons~\cite{Backenstoss, Rzehorz}, have been entirely speculative for $\bar{K}$.
 
 We therefore have investigated the correlations of  back-to-back coincident $\Lambda d$ and $\Lambda t$ pairs from the stopped $K^-$ reaction on $^4$He, and clearly identified the both $\Lambda d$ and $\Lambda t$ branches of the three-Nucleon Abosorption (3NA) and four-Nucleon Absorption (4NA) processes of $K^-$, 
 \begin{eqnarray}
 (K^- - {}^4\textrm{He})_{atomic} & \rightarrow & \Lambda + d +``n" :\textrm{3NA}, \label{eq:Ldeqn}\\
 (K^- - {}^4\textrm{He})_{atomic} & \rightarrow & \Lambda + t :\textrm{4NA},\label{eq:Lteqn}
 \end{eqnarray}
 for the first time, respectively, and further evaluated the per-stopped-$K^-$ fractions of these reaction branches. This Letter presents the successful observations of the direct three- and four-nucleon absorption processes of $\bar{K}$ meson at rest in the $\Lambda d n$ and $\Lambda t$ final states, respectively, from the stopped $K^-$ reaction in $^4$He in the KEK-PS E549. Among the report, the $\Lambda d$ part here discusses an advanced study including the spectrum normalization and the improved deuteron selection procedure from the previous report~\cite{TS2}, and the $\Lambda t$ part does descrive the first clear observation of the 4NA process with the per-stopped-$K^-$ formation rate of its $\Lambda t$ branch.

\section{Experimental method and low-level analysis procedures}\label{EXP}
 The experiment E549 was performed at the K5 beamline at the KEK 12 GeV proton synchrotron. The original aim of the experiment was to study kaonic tribaryon states with $^4$He($K^-_{stopped}$, $p/n$) reactions, and the basic analysis procedures and results of the proton and neutron emmission channels have been already published in Refs.~\cite{Sato} and~\cite{Yim}, respectively. Thus, we concentrate here to describe the essential parts related to the coincidence measurements of back-to-back $\Lambda d$ and $\Lambda t$ pairs.
 
 The experimental setup and all related devices are illustrated in Fig. 1 of Refs.~\cite{Sato,Yim}, and we follow the notations adopted there. The detailed specifications of devices and detectors are also found in the references. The incident 650 MeV/c $K^-$ beam was slowed down, and stopped on the $^4$He target~\cite{Sato2}. Then, final state particles from the stopped $K^-$ reaction on the target were measured by the Time-Of-Flight (TOF) method.  The secondary particles were detected by two horizontal- and two vertical arms, on which one neutral- or charged particle could be detected, respectively. Therefore, the maximum number of final state particles detected in coincidence was four. 
 
\begin{figure}[h!]
\includegraphics[scale=.60]{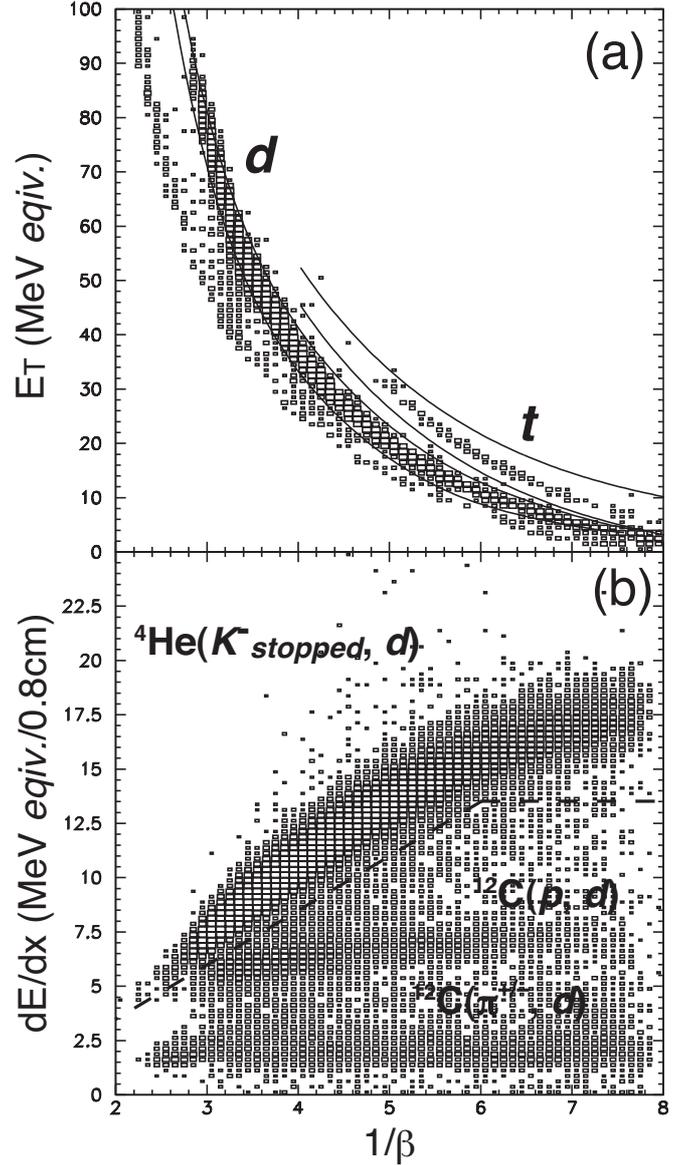}
\caption{(a) The measured correlation between $1/\beta$ (horizontal) and $E_T$ (vertical, given by the unit of MeV equivalent). The definitions of deuteron and triton events are overlaid by solid curves.  Protons and lighter particles are removed beforehand to focus on sparse heavy particles. \newline
(b) The correlation between $1/\beta$ (horizontal) and $dE/dx$ measured by Pstart (vertical) for deuteron events identified by $1/\beta$ and $E_T$, together with the adopted boundary between primary and secondary deuteron components drawn by a black dashed line. Secondary deutron components  from ${}^{12}$C$(\pi,d)$ and ${}^{12}$C$(p,d)$ reaction on Pstart, which appear as horizontal band structure with $dE/dx \sim 2$ and $\sim 6$, respectively, are removed by the second-stage deuteron selection.}\label{fig:fig1}
\end{figure}

\begin{figure}[h!]
\includegraphics[scale=.45]{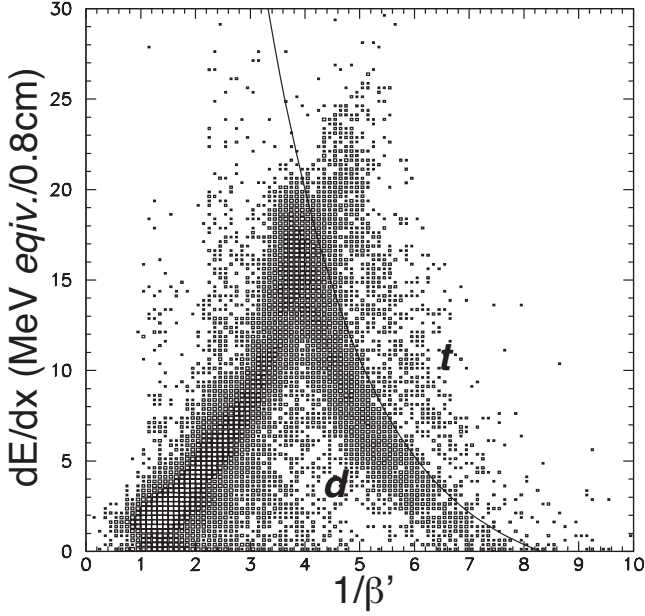}
\caption{The measured correlation between $1/\beta'$ (horizontal) and $dE/dx$ (vertical) on Pstart. The fast particles penetrating the Pstart appears as the left half of the intense wedge-shaped component with the possitive correlation, while slow deuterons stopped on Pstart form the right half. On the upper-right of the first wedge, triton component is observed as a weak second wedge. In the analysis, triton events are defined as those events above the border drawn by black real curve on the plot.}\label{fig:fig2}
\end{figure}

 Each horizontal arm consisted of a drift chamber (PDC), TOF start (Pstart) and stop (Pstop) counter walls, and Neutron Counters (NC), in order from the target center.  The TOF measurement was performed between Pstart and Pstop, and the measured TOF and particle track were converted together into the velocity $\beta$. The seven layers of NC, which were originally designed and installed as the neutron detectors, worked as the electromagnetic calorimeter for charged particles to measure the total energy $E_T$ with enough thickness to make $p$, $d$ and $t$ from the stopped $K^-$ reaction stop inside. The first-stage particle identification to select preliminary deuteron events and energetic tritons could reach Pstop, was then performed by extracting the corresponding band structures on the correlation between $E_T$ and $1/\beta$ as illustrated in Fig.~\ref{fig:fig1} (a). Requiring the consistency between $dE/dx$ measured on Pstart and $1/\beta$ in the second stage, secondary deuterons produced by $^{12}$C($\pi^{\pm}$,$d$) and $^{12}$C($p$,$d$) reactions on Pstart were excluded as demonstrated in Fig.~\ref{fig:fig1} (b), so that the deuteron events were qualified further compared to those in the previous report~\cite{TS2} where the first-stage identification was exclusively used.  Fig.~\ref{fig:fig2} shows the second triton identification by $dE/dx$ and velocity $\beta'$ exclusively defined by Pstart so as not to lose tritons could not reach Pstop. The final triton events were defined as a sum of these complemental two triton event sets. These two detector arms were symmetrically placed at the left and right sides of the target with their Pstop faces positioned at 1.8 m away from the center, so that $p$ and $d$ were measured with comparable r.m.s. resolutions of $\Delta(1/\beta)=0.02$. The configuration allowed the coincidence detection of ``back-to-back" $dp$ and $tp$ pairs emitted horizontally, and the reaction vertex, namely, the stopping point of $K^-$, was defined as the cross point of the incident $K^-$ and $d/t$ tracks.
 
 On the other hand, two vertical detector arms located on the top and bottom of the target, and each of them consisted of a drift chamber (VDC) and four layers of segmented scintillation counters with different thicknesses (TC) as drawn in Fig. 1 of Ref.~\cite{Yim}.  The identification of the incident charged particle was performed by the correlation between $dE/dx$ and $E$ measured in TC layers, and $\pi^{\pm}$ were successfully extracted, as described in detail in Ref.~\cite{Yim}. When a proton was detected in one of horizontal arms, the coincident pion velocity was calculated by adopting the decay vertex defined as the cross point of the pion (in VDC) and proton (in PDC) tracks, and TOF between the proton-fired Pstart and pion-fired TC with the correction for the proton flight time from the decay vertex to Pstart. Therefore, the velocity could be accurately evaluated without possible deviation due to the life time of $\Lambda$ or meta-stable kaon atomic orbits without proper start counters in these arms.
 
 These $p$ and $d$ on horizontal arms and $\pi^{\pm}$ on a vertical arm were corrected for energy loss adopting corresponding vertices and velocity 3-vectors, and their 4-momenta were reconstracted. $\Lambda$ hyperons were identified by means of invariant mass spectroscopy of $p$ and $\pi^{\pm}$ as shown in Fig. 2 (a) of Ref.~\cite{TS3}, and its momentum, $p_{\Lambda}$, was determined within $1\%$ scale discrepancy by the resolution rightly expected from $\Delta(1/\beta)=0.02$ for the whole interested momentum region.  The acceptance of the opening angle of two 3-momenta, $\theta_{\Lambda d/\Lambda t}$, was limited in the region, $\cos{\theta_{\Lambda d/\Lambda t}}< -0.7$, by the detector geometry. The resolution and accuracy of the opening angle and deuteron momentum, $p_d$, were demonstrated by simultaneously observed $^4_{\Lambda}\textrm{He} \rightarrow d + d$ non-mesonic two-body decay process~\cite{Outa2}.  Correlations between the $\Lambda$ and coincident $d/t$ were investigated in terms of these kinematic variables.

Only incident $K^-$ and an outgoing charged particle were required as the minimum-biased hardware trigger, and the coincidence events of back-to-back $p+d/t$ and $\pi$ in the vertical direction were extracted by offline analysis from the accumulated data in the three different periods. The first measurement was performed in June 2005 as the dedicated beam time for E549. The second and third ones were during October 2005 and December 2005, respectively, as the parasitic measurements of kaonic $^4$He X-ray precision spectroscopy, E570~\cite{Okada}. In total, $(2.5 \pm 0.4) \times10^8$ stopped $K^-$ on the $^4$He target were realized.

\section{Normalized $\Lambda d$ spectra in a $\Lambda d n$ final state}\label{3NA}

511 back-to-back $\Lambda d$ pairs were successfully reconstracted from 893 $pd$+$\pi^{\pm}$ events obtained by the updated deuteron selection procedure described in the previous section. 
The uncorrected $\Lambda d$ spectra, resolution and so on, are already discussed in detail in Ref.~\cite{TS2}. Here, we concentrate to the normalization procedures and features of the normalized spectrum from a $\Lambda d n$ final state.

The spectrum function $\phi(\vec{p}_{\Lambda},\vec{p}_d)$, which represents the per-stopped-$K^-$ probability of $\Lambda d$ emission with three-momenta $\vec{p}_{\Lambda}$ and $\vec{p}_d$ per phase volume, is related to the total number of stopped $K^-$, $N_{K^-_{stopped}}$, and detected number of $\Lambda d$ pairs in the volume through the measurement, $N_{\Lambda d}(\vec{p}_{\Lambda},\vec{p}_d) \cdot d^3\vec{p}_{\Lambda} d^3 \vec{p}_d$, as 
\begin{eqnarray} 
 &&N_{\Lambda d}(\vec{p}_{\Lambda},\vec{p}_d) \cdot d^3\vec{p}_{\Lambda} d^3 \vec{p}_d/\epsilon_{\Lambda d}(\vec{p}_{\Lambda},\vec{p}_d) \nonumber \\
  &=& N_{K^-_{stopped}} \cdot \phi(\vec{p}_{\Lambda},\vec{p}_d) \cdot d^3\vec{p}_{\Lambda} d^3 \vec{p}_d,
\end{eqnarray}
where $\epsilon_{\Lambda d}(\vec{p}_{\Lambda},\vec{p}_d)$ is the acceptance for the $\Lambda d$ coincidence detection as a function of $\vec{p}_{\Lambda}$ and $\vec{p}_d$. Due to the isotropic feature of the stopped $K^-$ reaction, the expression is reduced to 
\begin{eqnarray}
 &&N_{\Lambda d}(p_{\Lambda},p_d,\cos{\theta_{\Lambda d}}) / \epsilon_{\Lambda d}(p_{\Lambda},p_d,\cos{\theta_{\Lambda d}}) \nonumber \\
 &=& N_{K^-_{stopped}} \cdot \phi(p_{\Lambda},p_d,\cos{\theta_{\Lambda d}}).
\end{eqnarray}
Here, $\epsilon_{\Lambda d}(p_{\Lambda},p_d,\cos{\theta_{\Lambda d}})$ is exactly caluculated in the whole variable space considering the detector geometry, all important physical effects and analysis efficiencies to identify and reconstruct $\Lambda$ and $d$ with their three-momenta, so that the acceptance is generally applicable regardless of the physics processes to produce the pairs.
 
In addition, it is known that a significant fraction of stopped negative particles such as $K^-$ is trapped by metastable atomic orbits and goes through the free decay. Probability of the free decay, $b_{free}$, is accurately known as $3.5 \pm 0.5 \%$ per stopped $K^-$ on $^4$He~\cite{Outa1}, and thus, $N_{K^-_{stopped}}$ is derived from the simultaneously measured number of monochromatic $\mu^-$, $N_{K\mu2}$, from the delayed two-body free-decay of $K^-$ via $K^-_{atomic} \rightarrow \mu^- \bar{\nu}_{\mu}$, as 
\begin{equation}
N_{K^-_{stopped}} = \frac{1}{b_{free} \cdot Br(K^-\rightarrow \mu^- \bar{\nu}_{\mu}) }\frac{N_{K \mu 2}}{\epsilon_{\mu}},
\end{equation}
where $Br(K^-\rightarrow \mu^- \bar{\nu}_{\mu}) $ is the branching ratio of the $K\mu2$ decay, and $\epsilon_{\mu}$ is the calculated detection efficiency of the monochromatic $\mu^-$. Consequently, $\phi(p_{\Lambda},p_d,\cos{\theta_{\Lambda d}})$ is evaluated as 
 \begin{eqnarray}
& & \frac{\phi(p_{\Lambda}, p_d,\cos{\theta_{\Lambda d}})}{b_{free} \cdot Br(K^-\rightarrow \mu^- \bar{\nu}_{\mu})} \nonumber \\
&=& \frac{N_{\Lambda d}(p_{\Lambda}, p_d,\cos{\theta_{\Lambda d}} ) / \epsilon_{\Lambda d}(p_{\Lambda}, p_d,\cos{\theta_{\Lambda d}})}{N_{K \mu 2} / \epsilon_{\mu}}.
\end{eqnarray}
The reconstructed $\Lambda d$ events are sorted out into the three-dimensional frequency table $N_{\Lambda d}(p_{\Lambda}, p_d,\cos{\theta_{\Lambda d}})$, and the spectrum function $\phi$ was evaluated by bin-by-bin acceptance correction.  Then, one- or two dimensional distributions of physical quantities were computed from $\phi$. In order to avoid large errors arisen from bins with tiny count number and acceptance, we have investigated the spectra in a limitted phase space, $p_d \ge 550$ MeV/$c$, $p_{\Lambda} \ge 350$ MeV/$c$, and $\cos{\theta_{\Lambda d}} \le -0.8$. 
 
\begin{figure*}[h!]
\begin{center}
\includegraphics[scale=.45]{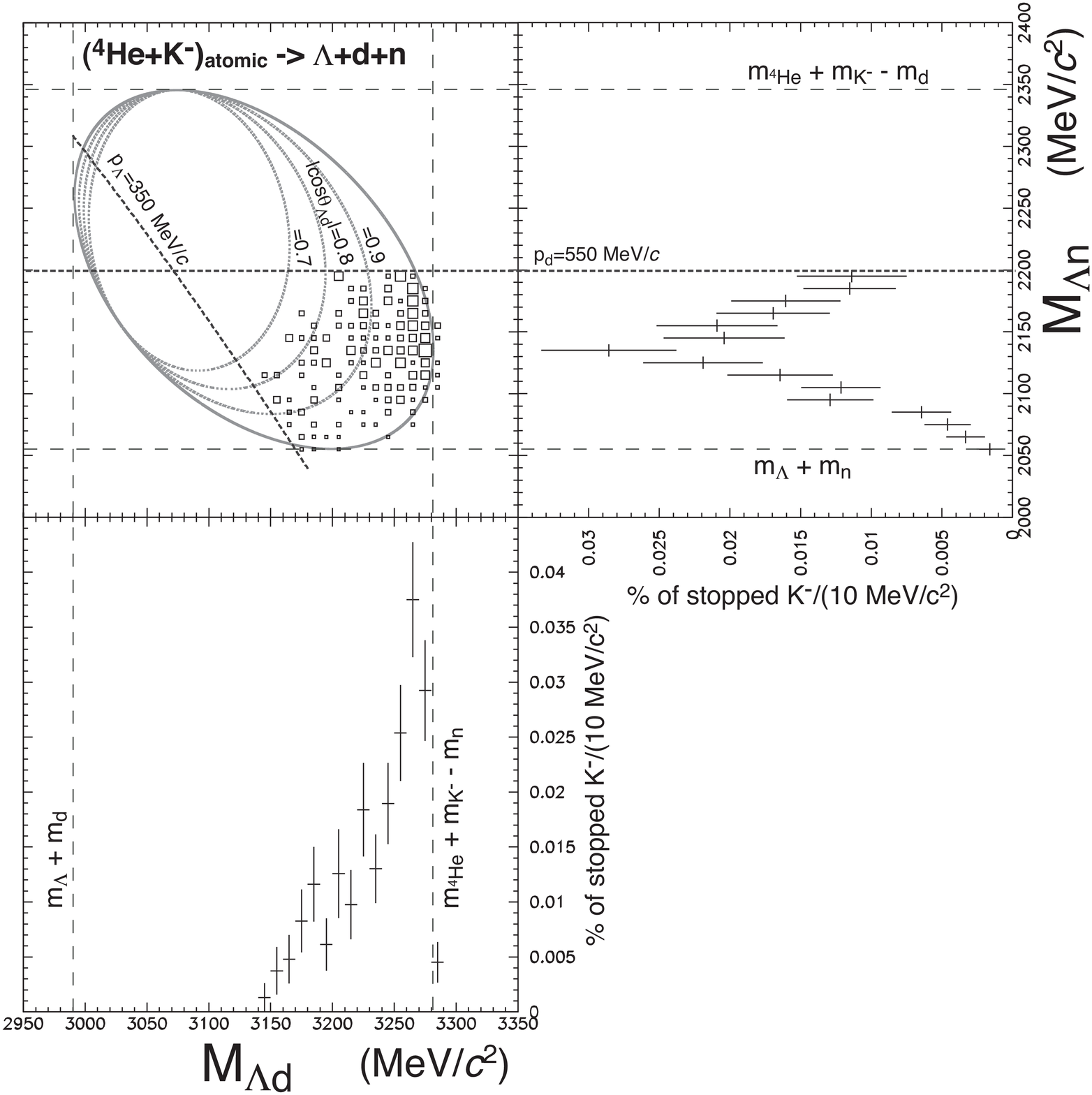}
\end{center}
\caption{The normalized scatter plot of $M_{\Lambda d}$ (horizontal) and $M_{\Lambda n}$ (vertical) of the $(^4\textrm{He}-K^-)_{atomic} \rightarrow \Lambda + d + n$ reaction, together with the projections onto each axis. On the plot, momentum thresholds of $p_{\Lambda} = 350$ MeV/$c$ and $p_d = 550$ MeV/$c$ are indicated by black dashed lines, and $|\cos{\theta_{\Lambda d}}|=const.$, and full kinematic limit, {\it i.e.} $|\cos{\theta_{\Lambda d}}|=1$, are drawn as gray real curves. The phase region visible here is limited to the region surrounded by $p_{\Lambda} = 350$ MeV/$c$, $p_d = 550$ MeV/$c$, and $-1 \le \cos{\theta_{\Lambda d}} \le -0.8$. The error bars on the projections represent statistical error only.}\label{fig:fig3}
\end{figure*}

The $\Lambda d n$ three-body final state is separated in terms of $^4$He($K^-_{stopped}, \Lambda d$) missing mass,
\begin{equation}
M_{N^*}=\sqrt{(m_{^4\rm{He}}+m_{K^-}-E_{\Lambda}-E_d)^2-(\vec{p}_{\Lambda}+\vec{p}_{d})^2},
\end{equation}
 which is actually the total mass of missing one-baryon system. Then, the $^4$He($K^-_{stopped}, d$) missing mass,
 \begin{equation}
 \sqrt{(m_{^4\rm{He}}+m_{K^-}-E_d)^2-{p_{d}}^2},
 \end{equation} 
 is kinematically equivalent to the total mass of the system consisted of $\Lambda$ and undetected $n$, $M_{\Lambda n}$, and hence the reaction dynamics can be investigated as possible nonuniformity in the three-body phase space of $(^4$He+$K^-)_{atomic} \rightarrow \Lambda+d+n$ reaction. 
 
 The normalized scatter plot of  $\Lambda d$ invariant mass $M_{\Lambda d}$ and $M_{\Lambda n}$ is shown in Fig.~\ref{fig:fig4}. As illustrated in the figure, a limited region out of the full phase space of the reaction is visible in our acceptance. At the right edge of the visible region, we observe a strong event cluster, which appears as a prominent peak structure in the projection onto $M_{\Lambda d}$ axis. As discussed in Ref.~\cite{TS2},  this corresponds to the 3NA process of stopped $K^-$ (\ref{eq:Ldeqn}). The per stooped $K^-$ branching fraction of the process, which is estimated by the integration of the spectrum above 3240 MeV/$c^2$, is 
 \begin{equation}
 Br(\Lambda d) = \bigl(0.9 \pm 0.1(stat) \pm 0.2 (syst)\bigr)\times 10^{-3},
 \end{equation} 
 where constant background of $0.05\times 10^{-3}$ per 10 MeV/$c^2$ is assumed, and the systematic error originates from dominant contribution of 14$\%$ relative uncertainty of $b_{free}$ and smaller uncertainties of the background and signal shape in the sensitive region and the distrubution of the 3NA contribution in the insensitive region.

\section{$\Lambda$ spectrum under triton coincidence condition}\label{4NA}
We identified 1519 $pt$+$\pi^{\pm}$ events, from which 464 back-to-back $\Lambda t$ pairs were finally reconstracted. The $\Lambda t$ events were contaminated by the $\Lambda d$ to a certain degree due to the miss-identification of slow $d$ as $t$ on Pstart. In contrast to the $\Lambda d$ case, no selection was applied on the possible missing particles, and hence the pairs were considered to originate mainly from $\Lambda t$ and $\Lambda \gamma (\Sigma^0) t $ non-mesonic final states. Contribution from mesonic final states, $\Lambda \pi^0 t$ and  $\Lambda \gamma (\Sigma^0) \pi^0 t$, was expected not to be dominant because of a high detection threshold of the triton momentum. 

\begin{figure}[h!]
\includegraphics[scale=.45]{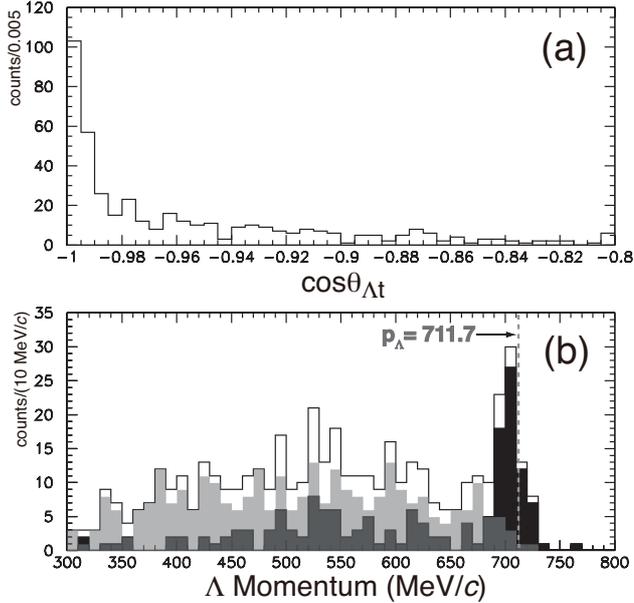}
\caption{(a) The distribution of $\cos{\theta_{\Lambda t}}$. The $\Lambda t$ events cluster below -0.99. (b) $\Lambda$ moentum spectra in coincidence with a backward $t$. The black-filled area corresponds to the events set with $\cos{\theta_{\Lambda t}} \le -0.99$, while the transparent gray is the complement, $i. e.$ $\cos{\theta_{\Lambda t}}>-0.99$. }\label{fig:fig4}
\end{figure}

The distribution of $\cos{\theta_{\Lambda t}}$ is given in Fig.\ref{fig:fig4} (a).  The events cluster below $\cos{\theta_{\Lambda t}}=-0.99$ exhibiting a remarkable back-to-back feature,  and the event rate decreases slowly with $\cos{\theta_{\Lambda t}}$ above. Fig.\ref{fig:fig4} (b) shows the $\Lambda$ momentum spectrum together with its classification by $\cos{\theta_{\Lambda t}}$.  On the spectrum, a prominent peak structure exists at $\sim705$ MeV/$c$, and it originates from the events with $\cos{\theta_{\Lambda t}} \le -0.99$ almost exclusively as is clearly demonstrated by the comparison of the classified spectra, implying its two-body nature. The Gaussian center and standard deviation of the observed peak are $704.7 \pm 1.4$ MeV/$c$ and $11.7 \pm 1.0$ MeV/$c$, respectively. The central position agree with the expected peak position of 711.7 MeV/$c$ for $(K^--^4\textrm{He})_{atomic} \rightarrow \Lambda + t$ reaction within $\sim$ 1$\%$ accuracy, and the standard deviation is consistent to the experimental momentum resolution of 12 MeV/$c$. Therefore, we now conclude that the observed monochromatic peak structure corresponds to the $\Lambda t$ branch of the direct 4NA process (\ref{eq:Lteqn}) beyond doubt.  The observed peak count $N_{\Lambda t}$ is $72 \pm 8 (stat) \pm 2 (syst)$, for which the systematic error originates from ambiguity of the background shape in the peak region. In the identical scheme descrived in the previous section for the $\Lambda d$ case, the count number is converted into the per-stopped-$K^-$ formation fraction, $Br(\Lambda t)$, as
 \begin{eqnarray}
Br(\Lambda t) &=& b_{free} \cdot Br(K^-\rightarrow \mu^- \bar{\nu}_{\mu}) \cdot \frac{N_{\Lambda t} / \epsilon_{\Lambda t}}{N_{K \mu 2} / \epsilon_{\mu}} \nonumber \\
&=& \bigl( 3.1 \pm 0.4(stat) \pm 0.5(syst)\bigr) \times 10^{-4},
\end{eqnarray}
where $\epsilon_{\Lambda t}$ is the detection efficiency of the process (\ref{eq:Lteqn}) calculated by GEANT3, taking all geometrical effects and analysis efficiency arisen from particle identification procedures of both  $\Lambda$ and $t$ into account. The systematic error is again dominated by the uncertainty of $b_{free}$.
 
 The continuum between $\sim 300$ MeV/$c$ and $\sim 700$ MeV/$c$ is considered to consist of the $\Sigma^0 t$ branch of the 4NA process, $ (K^- - {}^4\textrm{He})_{atomic} \rightarrow \Sigma^0 + t $ followed by $\Sigma^0 \rightarrow \Lambda \gamma$ ($480 \sim 680$ MeV/$c$ in $\Lambda$ momentum), and contaminant contribution of $\Lambda dn$/$\Sigma^0 d n$ 3-body final states.

\section{Discussion and conclusions}
Until now, any indication of the direct 3NA process of $\bar{K}$ meson has not been given anywhere.  Only one possibility was FINUDA experiment, where back-to-back correlated fast $\Lambda d$ pairs were observed as well in the stopped $K^-$ reaction on ${}^6$Li and ${}^{12}$C, and similar peak structure just below the $m_{\Lambda}+m_d$ mass threshold was observed especially in the ${}^6$Li case. However, the peak structure was assigned to the $\Lambda d$ decay of slow-moving $K^-ppn$ cluster, not to the direct 3NA process \cite{Piano1}. In our case, contributions from $\Lambda d n$ and $\Sigma^0 d n$ final states were separated from observed $\Lambda d$ pairs, and the normalized $\Lambda d$ invariant mass spectrum from a $\Lambda d n$ final state was presented in a limited region of the whole phase space. The observed peak just below the threshold was assigned to the $(^4\textrm{He}-K^-)_{atomic} \rightarrow \Lambda + d+``n"$ reaction with the branching fraction of $(0.9 \pm 0.1(stat) \pm 0.2(syst)) \times 10^{-3}$, as the first evidence of the 3NA process of $\bar{K}$ meson.

On the other hand, the existence of the direct 4NA process, $(^4\textrm{He}-K^-)_{atomic} \rightarrow \Lambda + t$, had been already indicated by Roosen {\it et al.}~\cite{Roosen}. In their result, however, all three event candidates could be attributed to the both $\Lambda t$ and $\Lambda d n$ final states, were assumed to be the $\Lambda t$ events, and the branching fraction of the hypothetical process was evaluated.  Long time later, FINUDA also observed back-to-back correlated $\Lambda t$ pairs, and they interpretted them as consequences of stopped $K^-$ captured by a $\alpha$-like cluster in the target nuclei, ${}^{6,7}$Li and ${}^9$Be. Very unfortunately, their source could be either a direct, or a more composite process which involves an intermediate bound system $K^-ppnn$, $K^-_{stop} \alpha \rightarrow (K^-ppnn) \rightarrow \Lambda t$, and a specific reaction mechanism could not be disentangled~\cite{Piano2}. Therefore, existence of the direct 4NA process of $\bar{K}$ meson is considered to be established now by this report, for the first time. The observed branching fraction $Br(\Lambda t)=(3.1 \pm 0.4(stat) \pm 0.5(syst)) \times 10^{-4}$ agrees well with the hypothetical value of $(3 \pm 2) \times 10^{-4}$~\cite{Roosen}.

Although the existence of 3NA and 4NA processes has been now established in the stopped $K^-$ reaction, the observed branching fractions, $\sim10^{-3}$ and $\sim 10^{-4}$ per stopped $K^-$, respectively, are significantly smaller than that of 2NA processes by one or two orders. Consequently, $YNN$ and $YNNN$ branches of these processes are not expected to account for the rather large discrepancy between the total fraction of the non-mesonic final states, and of the 2NA processes, which is recently observed dynamically as the non-mesonic but non-2NA spectrum strength in the $\Lambda N$ spectra with the branching fraction of $1\%$ order~\cite{FINUDA,TS3}. Therefore, the discrepant part must involve yet unkown exotic non-mesonic reaction branches, and further investigation of the part is still essential to achieve the full understanding of the $\bar{K}$-nuclear interaction.

\section{Acknowledgements}
We are grateful to the KEK staff members of the beam channel group, accelerator group, and electronics group, for support of the present experiment. We also owe much to T. Taniguchi and M. Sekimoto for their continuous contribution and advices for electronics. This work was supported by KEK, RIKEN, and Grant-in-Aid for Scientific Research (S) 14102005 of the Ministry of Education, Culture, Sports, Science and Technology of Japan.



\bibliographystyle{model1-num-names}
\bibliography{3NA-4NA}







\end{document}